# Deep Karaoke: Extracting Vocals from Musical Mixtures Using a Convolutional Deep Neural Network

Andrew J.R. Simpson [#1], Gerard Roma [#2], Mark D. Plumbley [#3]

[#] Centre for Vision, Speech and Signal Processing, University of Surrey
Guildford, UK
[1] andrew.simpson@surrey.ac.uk
[2] g.roma@surrey.ac.uk
[3] m.plumbley@surrey.ac.uk

*Abstract*—**Identification and extraction of singing voice from within musical mixtures is a key challenge in source separation and machine audition. Recently, deep neural networks (DNN) have been used to estimate 'ideal' binary masks for carefully controlled cocktail party speech separation problems. However, it is not yet known whether these methods are capable of generalizing to the discrimination of voice and non-voice in the context of musical mixtures. Here, we trained a convolutional DNN (of around a billion parameters) to provide probabilistic estimates of the ideal binary mask for separation of vocal sounds from real-world musical mixtures. We contrast our DNN results with more traditional linear methods. Our approach may be useful for automatic removal of vocal sounds from musical mixtures for 'karaoke' type applications.**

## I. INTRODUCTION

Much work in audio source separation has been inspired by the ability of human listeners to maintain separate auditory neural and perceptual representations of competing speech in 'cocktail party' listening scenarios [1]-[3]. A common engineering approach is to decompose a mixed audio signal, comprising two or more competing speech signals, into a spectrogram in order to assign each time-frequency element to the respective sources [4]-[6]. Hence, this form of source separation may be interpreted as a classification problem.

A benchmark for this approach is known as the 'ideal binary mask' and represents a performance ceiling on the approach by providing a fully-informed separation based on the spectrograms for each of the component source signals. Using the source spectrograms, each time-frequency element of the mixture spectrogram may be attributed to the source with the largest magnitude in the respective source spectrogram. This ideal binary mask may then be used to establish reference separation performance. In a recent approach to binary-mask based separation, the ideal binary mask was used to train a deep neural network (DNN) to directly estimate binary masks for new mixtures [6]. However, this approach was limited to a single context of two known speakers and a sample rate of only 4 kHz. Therefore, it is not yet known whether the approach is capable of generalizing to less well controlled scenarios featuring unknown voices and unknown background sounds. In particular, it is not known whether such a DNN architecture is capable of generalizing to the more demanding task of extracting unknown vocal sounds from within unknown music [7]-[9].

In this paper, we employed a diverse collection of real-world musical multi-track data produced and labelled (on a song-by-song basis) by music producers. We used 63 typical 'pop' songs in total, each featuring vocals of various kinds. For each multi-track song/mix, comprising a set of component 'stems' (vocals, bass, guitars, drums, etc), we pooled audio labeled as 'vocal' separately to all other audio (i.e., the accompanying instruments). We then obtained arbitrary mixtures for each song, simulating the process of mixing to produce 'mixes' for each song. Using the first 50 songs as training data, we trained a convolutional DNN to predict the ideal binary masks for separating the respective vocal and non-vocal signals for each song. For reference, we also trained an equivalent linear method (convolutional non-negative matrix factorization - NMF) of similar scale. We then tested the respective models on mixes of new songs featuring different musical arrangements, different singing and different production. From both models we obtained probabilistic estimates of the ideal binary mask and analyzed the resulting separation quality using objective source separation quality metrics. These results demonstrate that a convolutional DNN approach is capable of generalizing voice separation, learned in a musical context, to new musical contexts. We also illustrate the capability of the probabilistic convolutional approach [6] to be optimized for different priorities of separation quality according to the statistical interpretation employed. In particular, we highlight the differences in performance for the two respective architectures in the context of the trade-off between artefacts and separation.

## II. METHOD

We consider a typical simulated ensemble musical performance scenario featuring a variety of musical contexts and a variety of vocal performances. In each context, which we refer to as 'a song', there are a multitude of musical accompaniment signals and at least one (often more) vocal signals. The various signals are mixed together (arbitrarily) and the resulting mixture is refered to as 'a mix'. The engineering problem is to automatically separate all vocal signals from the concurrent accompaniment signals. We used 63 fully produced songs, taken from the MedleyDB database [10]. The average duration of the songs was 3.7 minutes (standard deviation (STD): ±2.7 mins). The average number of accompanying sources (stems) was 7.2 (STD: ±6.6 sources) and the average number of vocal sources was 1.8 (STD: ±0.8 sources).

For each song, the source signals were classified as either vocal or non-vocal (according to the labels assigned by the music producers). Vocal sounds included both male and female singing voice and spoken voice ('rap'). Non-vocal sounds included accompanying instruments (drums, bass, guitars, piano, etc). Source sounds were studio recorded and featured relatively little interference from other sources. All source sounds were then peak normalized before being linearly summed into either a vocal mixture or a non-vocal mixture respectively. The two separate (vocal / non-vocal) mixtures were then peak normalized and linearly summed to provide a complete mixture (i.e., a 'final production mix'). This provided for a mixture that resembled a mix that might be produced by a human mixing engineer [11]. All sources and mixtures were monaural (i.e., we did not employ any stereo processing).

All signals were sampled at a rate of 44.1 kHz. The respective source (vocal / non-vocal) and mixture signals were transformed into spectrograms using the short-time Fourier transform (STFT) with window size of 2048 samples, overlap interval of 512 samples and a Hanning window. This provided spectrograms with 1025 frequency bins. The phase component of each spectrogram was removed and retained for later use in inversion. From the source spectrograms a binary mask was computed where each element of the mask was determined by comparing the magnitudes of the corresponding elements of the source (vocal / non-vocal) spectrograms and assigning the mask a '1' when the vocal spectrogram had greater magnitude and '0' otherwise.

The first 50 songs (taken in arbitrary order) were used as training data and the final 13 songs were used as test data. The magnitude-only mixture spectrograms computed from the first 50 songs and the respective ideal binary masks were used as training data. Note, phase was not used in training the model.

For the training data, the mixture spectrogram and the corresponding source spectrograms were cut up into corresponding windows of 20 samples (in time). The windows shifted at intervals of 60 samples (i.e., there was no overlap). Thus, for every 20-sample window, for training the models there was a mixture spectrogram matrix of size 1025x20 (frequency bins x time) samples and an ideal binary mask matrix of the same size. From the 50 songs designated as training data, this gave approximately 15,000 training examples. For the testing stage, the spectrograms for the remaining 13 songs were cut up with overlap intervals of 1 sample (which would ultimately be applied in an overlaping convolutional output stage). Prior to windowing, all spectrogram data was normalized to unit scale.

*Deep Neural Network.* We used a feed-forward DNN of size 20500x20500x20500 units (1025 x 20 = 20500). Each spectrogram window of size 1025 x 20 was unpacked into a vector of length 20500. The DNN was configured such that the input layer was the mixture spectrogram (20500 samples). The DNN was trained to synthesize the ideal binary mask at its output layer. The DNN employed the biased-sigmoid activation function [12] throughout with zero bias for the output layer. The DNN was trained using 100 full iterations of stochastic gradient descent (SGD). Each iteration of SGD featured a full sweep of the training data. Dropout was not used in training. After training, the model was used as a feed-forward probabilistic device.

*Probabilistic Binary Mask.* In the testing stage, there was an overlap interval of 1 sample. This means that the test data described the mixture spectrogram in terms of a sliding window and the output of the model described predictions of the ideal binary mask in the same sliding window format. The output layer of the DNN was sigmoidal and hence we may interpret these predictions in terms of the logistic function. Therefore, because of the sliding window, this procedure resulted in a distribution (size 20) of predictions for each time-frequency element of the mixture spectrogram [6]. We chose to summarize this distribution by taking the mean and we evaluate the result in terms of an empirical confidence estimate, separately for each source, as follows: For each time-frequency element, of each source, we computed the mean prediction and applied a confidence threshold ($\alpha$);

$$M^V{}_{t,f} = \begin{cases} 1 & for \quad \frac{1}{T}\sum_{i=0}^{T} S_{t+i,f} > \alpha \\ 0 & for \quad \frac{1}{T}\sum_{i=0}^{T} S_{t+i,f} \leq \alpha \end{cases} \quad (1)$$

where $M^V$ refers to the binary mask for the vocal source, $T$ refers to the window size (20), $t$ is the time index, $i$ is the window index and $f$ is the frequency (bin) index into the estimated mask ($S$). The corresponding (but independent) binary mask for the non-vocal source ($M^{NV}$) is computed as follows;

$$M^{NV}{}_{t,f} = \begin{cases} 1 & for \quad \frac{1}{T}\sum_{i=0}^{T} S_{t+i,f} < (1-\alpha) \\ 0 & for \quad \frac{1}{T}\sum_{i=0}^{T} S_{t+i,f} \geq (1-\alpha) \end{cases} \quad (2)$$

Thus, by adjustment of $\alpha$, masks at different levels of confidence could be constructed for both sources.

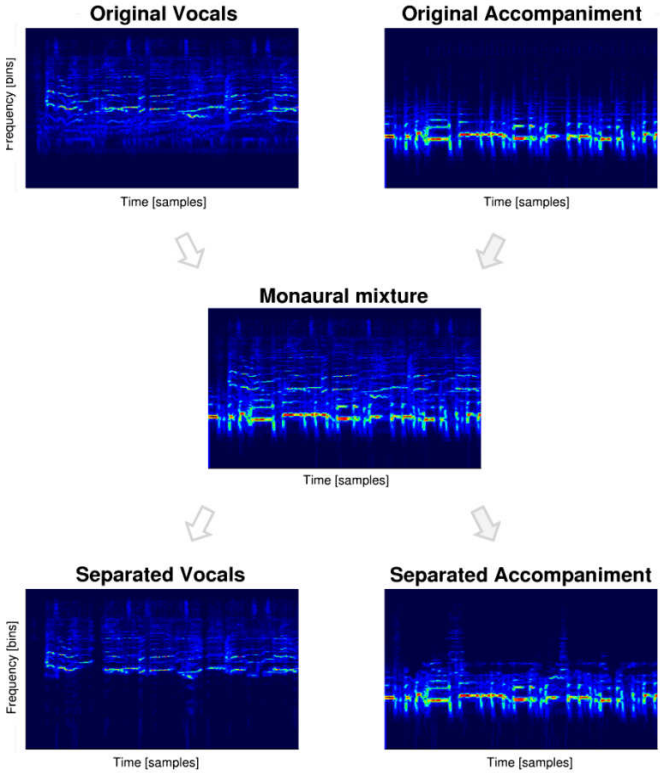

**Fig. 1. Separation of vocal sounds from musical mixtures using a probabilistic convolutional deep neural network.** The upper pair of spectrograms plot a ~1.5-second excerpt from a typical song taken from the test data set, illustrating the original monaural audio for the voice and non-voice (i.e., accompaniment) sources respectively. The middle spectrogram plots the monaural mixture (i.e., the ensemble music). The lower pair of spectrograms plot the respective separated channels ($\alpha = 0.5$). Note the frequency axis represents the range 0 – 22 kHz on a logarithmic axis.

*Non-negative matrix factorization.* For comparison to the DNN approach, an equivalent non-negative matrix factorization (NMF) based approach was implemented using the same training and test data (as described above). We used the same unpacking strategy, which has been tested before for NMF-based separation of speech and music [13]. The spectrograms of the training data were sampled and unpacked analogously to the DNN approach, resulting in a large (220500x15000) matrix that was then decomposed using the traditional multiplicative updates algorithm with KL divergence [14]. This means that for this training matrix $V$, $V = WH$, where we set the number of basis vectors (columns of $W$) and the respective activations (rows of $H$) to 1500. We performed this training stage for both vocal and non-vocal mixtures, and kept the two basis vectors matrices $W_v$ and $W_{nv}$. For the testing stage, we concatenated both matrices and initialized a corresponding $H_u$ matrix randomly, so that for each unpacked spectrogram, $V_u$, of the set of test songs, $V_u = [W_v \ W_{nv}] \ H_u$. We then ran the same multiplicative updates algorithm but keeping the composite $W_u$ matrix fixed [13], and updating $H_u$. The test spectrogram was then re-composed for either vocal ($V_v = W_v H_v$) or non-vocal ($V_{nv} = W_{nv} H_{nv}$) vectors, and used to define a soft mask via the element-wise division $S_v = V_v / (V_v + V_{nv})$. The matrix was then packed back to the original spectrogram size by averaging the consecutive frames of the soft mask. This allowed us to define an equivalent $\alpha$ parameter (as used in the DNN approach) so that the binary mask $B_v = 1$ when $S_v > \alpha$, 0 when $S_v <= \alpha$ and analogously for $B_{nv}$ and $S_{nv}$.

Finally, the respective masks were resolved by multiplication with the original (complex) mixture spectrogram and the resulting masked spectrograms were inverted with a standard overlap-and-add procedure. Separation quality (for the test data) was measured using the BSS-EVAL toolbox [15] and is quantified in terms of signal-to-distortion ratio (SDR), signal-to-artefact ratio (SAR) and signal-to-interference ratio (SIR). Separation quality was assessed at different confidence levels by setting different values of $\alpha$.

### III. RESULTS

Fig. 1 plots spectrograms illustrating the stages of mixture and separation for a brief excerpt (~1.5 seconds) from a randomly chosen test song separated using the DNN (at $\alpha = 0.5$). The spectrograms for the source vocal and non-vocal signals are shown at the top. The middle panel plots the mixture spectrogram, illustrating the difficulty of the problem (even for an ideal binary mask). At the bottom of Fig. 1 are plotted spectrograms representing the separated audio for the vocal and non-vocal signals respectively (DNN, $\alpha = 0.5$).

The various objective source separation quality metrics (SDR/SIR/SAR) were computed for the separated sources, as estimated with each model, as a function of $\alpha$. The same measures were also computed for the ideal binary mask. Fig 2 plots a summary of the respective measures. For each measure, and for each separation context (DNN/ideal binary mask/NMF), Fig. 2a plots the mean across-song performance computed by first averaging the measures across vocal/non-vocal sources. Fig. 2b plots the across-song average for the vocal sources only and Fig. 2c plots the same for the non-vocal (accompaniment) sources only. Shaded areas and error bars represent 95% confidence intervals. The results for the DNN and NMF (as a function of $\alpha$) feature similar functions illustrating the trade-off between the various parameters as statistical confidence is adjusted. Both models provide similar intersection points and there is some evidence of performance advantage for the DNN. However, the slopes and shapes of the functions are qualitatively different. In particular, the DNN functions for SAR and SIR more closely resemble 'ideal' sigmoid functions. In this context, SAR and SIR may be interpreted as energetic equivalents of positive hit rate (SIR) and false positive rate (SAR). Hence, if these slopes are interpreted as being analogous to cumulative density functions (indexed using $\alpha$), then the DNN results might be interpreted as demonstrating a wider probability function that is closer to normally distributed. However, although these plots provide insight into the mapping of probability to performance, they do not provide a very interpretable comparison of the models. In particular, the plots do not allow us to interpret performance in like terms with respect to the critical trade-off between artefacts and separation.

In order to provide a like-for-like comparison, Fig. 3a plots mean SAR as a function of mean SIR for both models (taken from Fig. 2a) for the useable range of $0.1 < \alpha < 0.9$. Fig. 3b plots the same for the functions of Fig. 2b and Fig. 3c plots the same for the functions of Fig. 2c. Overall (Fig. 3a), the DNN provides ~3dB better SAR performance for a given SIR index. This advantage is mostly explained by the ~5dB advantage for the vocal sources (Fig. 3b) and only a small advantage is evident for the non-vocal signals (Fig. 3c).

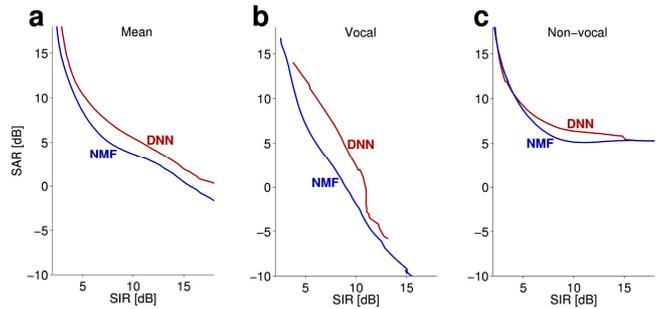

**Fig. 3. Trade-off: interference versus artefacts.** Illustrating the like-for-like performance of the respective models (DNN: red, NMF: blue); mean SAR as a function of mean SIR (taken from Fig. 2) for the useable range of $0.1 < \alpha < 0.9$. **a** plots the across-source, across-song mean (Fig. 2a), **b** plots the across-song mean for vocals (Fig. 2b) and **c** plots the across-song mean for non-vocals (Fig. 2c).

## IV. DISCUSSION AND CONCLUSION

We have demonstrated that a convolutional deep neural network is capable of separating vocal sounds from within typical musical mixtures. Our convolutional DNN is of nearly a billion parameters and was trained with relatively little data (and relatively few iterations of SGD). We have contrasted this performance with a like-for-like (suitably scaled) NMF approach, in the context of a trade-off between artefact and separation quality, indexed via confidence in the statistical predictions made.

The main advantage of the DNN appears to be in its general learning of what 'vocal' sounds are (Fig. 3b). Since the NMF approach is limited to linear factorization, we may at least partly attribute the advantage of the (nonlinear) DNN to abstract learning via demodulation [12]. The DNN appears to have biased it's learning towards making good predictions via correct positive identification of the vocal sounds.

Both methods feature the largest known parameterizations for this particular problem and, to some extent, both methods may be considered 'deep' [6], [12], [16]; both featured demodulated (magnitude) spectrograms produced using STFT and re-synthesis via inverse STFT. We also note that the relatively small amount of data employed in training the DNN may have been offset by the fact that the spectrograms were sampled using a Hanning window, hence minimizing aliasing/distortion in the training data that may otherwise have resulted in over-fitting [17] (and see [18]).

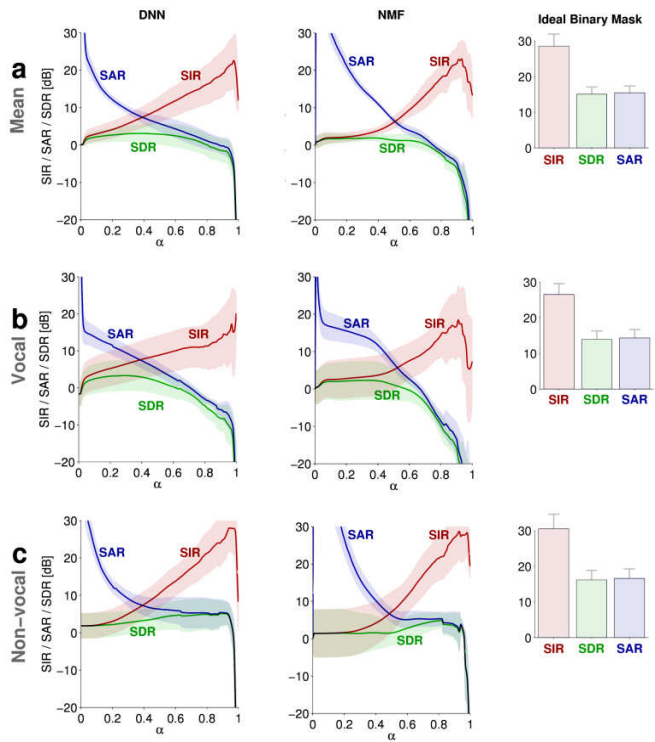

**Fig. 2. Separation quality as a function of $\alpha$: DNN versus NMF versus ideal binary mask.** The left hand column plots results obtained with the convolutional DNN, the central column plots benchmark results obtained using the ideal binary mask and the right hand column plots results obtained using the convolutional NMF approach. Source separation quality is evaluated in terms of signal-to-distortion ratio (SIR, red), signal-to-interference (SDR, green), signal-to-artefact ration (SAR, blue), computed from the entire duration of the 14 test songs using the BSS-EVAL toolkit [15]. **a** plots across-song mean, computed from across-source mean separation quality measures (i.e., computed across voice/non-voice sources). **b** plots mean quality measures for the vocal sources only and **c** plots the same for the non-vocal sources only. Shaded areas and error bars represent 95% confidence intervals.

## ACKNOWLEDGMENT

AJRS, GR and MDP were supported by grant EP/L027119/1 from the UK Engineering and Physical Sciences Research Council (EPSRC).